\documentclass[twocolumn,aps,prl,epsf,showpacs]{revtex4-1}

\usepackage{amsmath} 
\usepackage{amsfonts} 
\usepackage{epsfig} 
\usepackage{epsf} 
\usepackage{array}
\usepackage{color}
\begin{document}

\title{Unconventional proximity effect and inverse spin-switch behavior in a model manganite-cuprate-manganite trilayer system} 
\author{Juan Salafranca$^{1,2}$ and Satoshi Okamoto$^2$}
\affiliation{$^1$Department of Physics and Astronomy, The University of Tennessee, Knoxville, Tennessee 37996, USA\\
$^2$Materials Science and Technology Division, Oak Ridge National Laboratory, Oak Ridge, Tennessee 37831, USA}

\begin{abstract}
The proximity effect in a model manganite-cuprate system is investigated theoretically. 
We consider a situation in which spin-polarized electrons in manganite layers antiferromagnetically 
couple with electrons in cuprate layers as observed experimentally. 
The effect of the interfacial magnetic coupling is found to be much 
stronger than the injection of spin-polarized electrons into the cuprate region. 
As a result, the superconducting transition temperature depends on the thickness of the cuprate layer significantly. 
Since the magnetic coupling creates {\em negative} spin polarization, 
an applied magnetic field and the negative polarization compete resulting in the inverse spin-switch behavior 
where the superconducting transition temperature is increased by applying a magnetic field.  
\end{abstract}
\date{today}
\maketitle

Transition-metal oxides have been providing intriguing phenomena due to strong electron-electron or electron-lattice interactions, 
such as high-$T_c$ superconductivity (SC) in cuprates and the colossal magnetoresistance effect of manganites \cite{Imada98}.
Recent developments in fabricating atomically controlled heterostructures comprised of different transition-metal oxides 
allows us to explore further exotic phenomena that are not realized in bulk systems \cite{Izumi01,Ohtomo02,Okamoto04}.
Heterostructures involving cuprates and manganites have attracted much attention
because of the competition between SC and nearly full spin polarization, 
and their potential application as spintronic devices \cite{Sefrioui03,Varela03,Przyslupski04,Pena05,Chakhalian06,Chakhalian07}. 

This growing interest has rendered manganite-cuprate heterostructures 
the paradigmatic example of multilayers composed of oxides with competing ordered states. 
The electronic structure near the interface presents remarkable differences as compared to the bulk. 
The electronic charge is redistributed \cite{Yunoki07,Chakhalian07}, and 
orbital reconstruction takes place \cite{Chakhalian07}. 
As a consequence, spin polarization is induced in the cuprate side \cite{Chakhalian06}. 
The detailed characterization of the interface has been complemented by systematic studies of the collective properties. 

It has been revealed that the SC critical temperature $T_c$ of YBa$_2$Cu$_3$O$_{7-\delta}$ (YBCO) 
is more strongly suppressed when combined in superlattices with 
ferromagnetic (FM) La$_{1-x}$Ca$_x$MnO$_3$ (LCMO) \cite{Sefrioui03} than
with a nonmagnetic cuprate \cite{Varela99}. 
Even 5 unit-cell (u.c.) thick YBCO can become nonsuperconducting depending on the thickness of the LCMO layers . 
Since the $c$-axis coherence length in YBCO is about 1 u.c., 
these experiments indicate the existence of an unconventional
proximity effect.

More recently, a surprisingly robust inverse spin-switch effect (ISSE) 
was discovered in LCMO/YBCO/LCMO trilayer systems \cite{Nemes08,Dybko09}.   
In sharp contrast to the conventional exclusion between SC and magnetism \cite{review}, 
SC is favored by parallel alignment of the magnetization in the FM layers under an applied magnetic field. 
Changes in $T_c$ in these systems were found to be as high as 1.6~K \cite{Dybko09}. 
A similar ISSE has also been reported in conventional 
ferromagnet/superconductor heterostructures. 
However, the changes in $T_c$ were much smaller ($\sim$ 10 mK)
\cite{Yu06,Singh07},  
and these are most likely due to stray fields \cite{Steiner06}. 
A large band splitting between the majority and minority electrons could have such an effect \cite{Montiel09}. 
 But, the ISSE is shown to disappear when the minority band is above the Fermi level, 
and the mechanism of Ref.~\cite{Montiel09} cannot account for the ISSE in manganite-based heterostructures.

These experimental findings indicate that the interfacial phenomena in manganite-cuprate systems 
lie far outside of conventional theoretical models 
which merely consider the transfer of electrons at interfaces \cite{Yunoki07,Montiel09}. 
Identifying the proper interfacial interactions and providing physical pictures is, therefore, desired 
not only for understanding the experimental results but also 
for their potential device applications. 

-manganite trilayers. 
As the key ingredient, we consider the antiferromagnetic (AF) coupling at the interface confirmed experimentally \cite{Chakhalian06}. 
The AF coupling induces the negative spin polarization inside the cuprate region and 
influences SC more strongly than injecting spin-polarized quasiparticles without the coupling. 
We found that $T_c$ of such systems is drastically suppressed when 
the thickness of the cuprate layer is reduced, in accordance with the experimental reports. 
The balance between the effective field due to the interfacial AF coupling and external applied fields 
naturally explains the ISSE. 
The model also reproduces the characteristic length scales observed experimentally \cite{Sefrioui03} 
and the exponential decay of the ISSE with cuprate-layer thickness \cite{Nemes08}. 
Our work quantitatively links the details of the coupling at the manganite-cuprate interfaces 
and superconducting properties of the multilayers, 
and offers a coherent picture to understand most experimental results reported 
in manganite-cuprate heterostructures.

\begin{figure}[tbp] 
\includegraphics[width=0.9\columnwidth,clip]{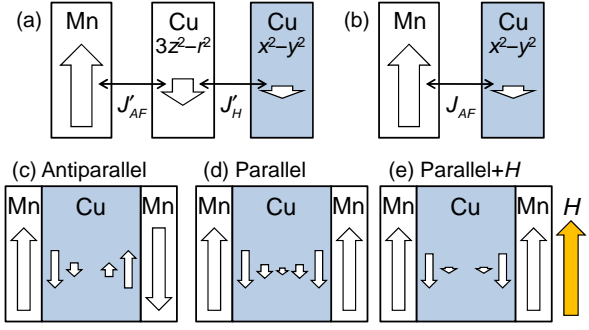} 
\caption{ Model. 
(a) Realistic manganite-cuprate interface and (b) simplified interface. 
Manganite conduction electrons and cuprate $3z^2-r^2$ electrons are coupled via the antiferromagnetic exchange $J'_{AF}$, 
while cuprate $3z^2-r^2$ and $x^2-y^2$ electrons are coupled ferromagnetically via the Hund coupling $J'_H$. 
Since the Hund coupling is larger than $J'_{AF}$, 
the magnetic interaction can be approximated as a direct coupling $J_{AF}$ as shown in (b).  
Trilayers with (c) an antiparallel configuration and (d) [(e)] a parallel configuration 
(with an applied magnetic field $H$). 
$H$ mostly affects the tails of the induced moment.} 
\label{fig:model} 
\end{figure}

{\em Model.}---
We consider a [001] trilayer system in which 
manganites and cuprates are stacked along the $z$ direction. 
The manganite region is described by the double-exchange model $H_{DE}$ 
with single conduction band for simplicity. 
The cuprate region is described by the BCS model $H_{BCS}$, with the
pairing involving Cu $d_{x^2-y^2}$ orbitals.  
At a manganite-cuprate interface, Cu $d_{3z^2-r^2}$ orbitals become
electrically active due to  the ``orbital reconstruction'' \cite{Chakhalian07,Yang09,Okamoto10}. 
The $d_{3z^2-r^2}$ orbital hybridizes with manganese $e_g$ orbitals, 
and the resulting AF coupling between them is $J'_{AF}$. 
Further, the $d_{3z^2-r^2}$ orbital and the Cu $d_{x^2-y^2}$ orbital on the same Cu site are expected to be coupled ferromagnetically 
via the Hund coupling $J_H'$ [see Fig.~\ref{fig:model}(a)]. 
Since $J_H' \gg J'_{AF}$, the Cu $d_{x^2-y^2}$ spin is expected to be slaved to the Cu $d_{3z^2-r^2}$ spin. 
Such a situation is modeled by a direct AF coupling $J_{AF}$ between a manganese spin and a Cu $d_{x^2-y^2}$ spin as shown in Fig.~\ref{fig:model}(b). 
The coupling constant $J_{AF}$ is the same order of magnitude as $J'_{AF}$ but somewhat reduced. 
Thus, the Hamiltonian of our model trilayer system is given by $H_{FSF} = \sum_{\alpha=R,L} \{ H_{DE,\alpha} + H_{int,\alpha} \} + H_{BCS} $ with 
\begin{eqnarray}
H_{DE,\alpha} \!\!&=&\!\! - \, t \!\!\!\! \sum_{\langle ij \rangle_{\in \alpha}, \sigma} \!\! \Bigl(c_{i \sigma}^\dag c_{j \sigma} + h.c. \Bigr) 
- J_H \sum_{i \in \alpha} \vec s^{\, Mn}_i \cdot \vec S_i, \,\, \,\, \\
H_{BCS} \!\!&=&\!\! \delta \varepsilon \!\!\! \sum_{i \in {\rm Cu}, \sigma} \!\!\! d_{i \sigma}^\dag d_{i \sigma} 
- \!\!\!\! \sum_{\langle ij \rangle_{\in {\rm Cu}}, \sigma} \!\! 
\Bigl(t_{ij} d_{i \sigma}^\dag d_{j \sigma} + h.c. \Bigr)  \,\, \,\, \nonumber\\
&& \hspace{-3.5em} + g \!\!\!\! \sum_{i \in {\rm Cu}, \tau = x,y} \!\!\!\! 
\Bigl( d_{i \uparrow}^\dag  d_{i+ \hat \tau \downarrow}^\dag 
d_{i + \hat \tau \uparrow}d_{i \downarrow}
+ d_{i \downarrow}^\dag d_{i + \hat \tau \uparrow}^\dag 
d_{i + \hat \tau \downarrow} d_{i \uparrow} \Bigr), \,\, \,\, \\
H_{int,\alpha} \!\!&=&\!\! \sum_{\langle i j \rangle'} \Bigl\{ \Bigl(v_\alpha c_{i \sigma}^\dag d_{j \sigma} + h.c. \Bigr) 
+ J_{AF} \vec s^{\, Mn}_i \cdot \vec s^{\, Cu}_j \Bigr\}. 
\end{eqnarray}
Here, $\alpha$=$R$ or $L$ indicates the manganite layer on the right or left, $J_H$ the Hund coupling in manganite regions, 
$c(d)$ is an electron annihilation operator in a manganite (cuprate) region, and 
$\vec s^{\, Mn}_i = \frac{1}{2} \sum_{\sigma \sigma'} c_{i \sigma}^\dag \vec \tau_{\sigma \sigma'} c_{i \sigma'}$, 
$\vec s^{\, Cu}_i = \frac{1}{2} \sum_{\sigma \sigma'} d_{i \sigma}^\dag \vec \tau_{\sigma \sigma'} d_{i \sigma'}$ with $\vec \tau$ the Pauli matrices, 
and $\vec S_i$ a localized $t_{2g}$ spin in the manganite regions ($|S_i|$=1). 
The transfer intensity in a manganite (cuprate) region is given by $t (t_{ij})$. 
The hybridization strength at a manganite-cuprate interface is given by $v_\alpha$ with 
$\langle i j \rangle'$ in $H_{int,\alpha}$ indicating the summation constrained for nearest-neighbor bonds across the interface. 
$g$ is the pairing coupling constant, 
$\hat x (\hat y)$ is the unit vector along the $x(y)$ direction, 
and $\delta \varepsilon$ indicates the band mismatch between the manganite and cuprate regions. 
Considering the experimental set up in Refs.~\cite{Nemes08,Dybko09}, 
i.e., the external magnetic field being applied in in-plane directions, 
the effect of the magnetic field $H$ is introduced by the Zeemann term 
$-\sum_i g_e s_i^z H$ with $g_e=2$, and $i$ running in both Cu and Mn regions. 

Since we are interested in the superconducting properties,
we focus on the temperature range far below the FM Curie temperature of manganites. 
$t_{2g}$ spins in manganite regions are then treated as classical degrees of freedom 
with parallel (P) or antiparallel (AP) spin alignment between the $R$ and $L$ manganite layers. 
We then apply the Bogoliubov-Hartree-Fock approximation by introducing
the order parameters 
$n^{Mn}_{i \sigma}$=$\langle c_{i \sigma}^\dag c_{i \sigma}\rangle, n^{Cu}_{i \sigma}=\langle d_{i \sigma}^\dag d_{i \sigma}\rangle$, and 
$\Delta_{i_z} = \langle d_{i \uparrow} d_{i+\hat x \downarrow} \rangle = - \langle d_{i \uparrow} d_{i+\hat y \downarrow} \rangle$
($d$-wave symmetry). 
These order parameters are determined by solving the self-consistent equations numerically. 

Each system in our simulations is characterized by the thickness of
the cuprate middle layer $N$, with the lattice constant taken to be unity. 
The total number of unit cells along $z$ in the manganite layers is made equal to $N$ to reduce the parameter space. 
In-plane hopping in the Cu region and isotropic hopping in the Mn regions 
are of the same order of magnitude and are larger than hopping along $z$ in the Cu region and across the interface. 
$J_{AF}$ and $g$ are derived from the superexchange-type processes. 
The effect of $J_{AF} (g)$ should be enhanced (suppressed) by the $e$-$e$ repulsion, 
although it is not included explicitly. 
Therefore, we use the following parameters: 
$v_R=v_L= 0.2t$, $J_{AF}= 0.6t$, 
$J_H= 10t$, $g=0.2t$, 
$t_{ij} = t \, (0.2t)$ for a nearest-neighbor transfer along the $xy
\, (z)$ direction,  
$\delta \varepsilon= 1.2t$, and 
the mean carrier density in the bulk cuprate (manganite) 0.8 (0.7). 
The bulk chemical potential for the cuprate region is about $1.5t$ lower 
than that in the manganite region. 
This results in a small electron doping of the cuprate region 
in the multilayer model systems \cite{Yunoki07}. 

As YBCO samples are always in the nearly optimal or underdoped sides of the phase diagram, 
these doping effects will enhance the suppression of $T_c$ in thin heterostructures, 
bringing the system closer to the Mott insulating phase.  
Enhanced phase fluctuations in thin heterostructures should also reduce $T_c$. 
But these effects are not included in the present treatment. 
The penetration depth of the SC order parameter in the 
FM region is extremely small due to the high spin polarization, 
and the $c$-axis SC coherence length is short in YBCO, 
therefore, results for trilayers in the P configuration with 
$N>2$ are representative of experiments in multilayers.

A larger $J_{AF}$ implies that the AF coupling 
has a larger effect suppressing the superconducting order,
and a larger $t_{ij \parallel z}$ reduces $T_c$ further, 
but our numerical results do not depend on the choice of these parameters in a significant way. 
For different system sizes, $\Delta$ is calculated as a function of
temperature and the magnetic field. $T_c$ is then defined as 
$\Delta (T_c) =10^{-2}\Delta(0)$. 
The numerical error is smaller than the point size presented below. 

\begin{figure}[tbp] 
\includegraphics[width=0.8\columnwidth,clip]{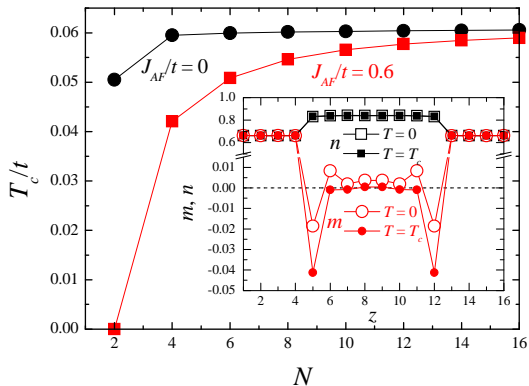} 
\caption{$T_c$ as a function of the Cu layer thickness $N$ in the P configuration. 
  $J_{AF}=0.6 t$ reduces $T_c$ by inducing the spin polarization in the Cu region, while 
  $J_{AF}=0$ shows the weak $N$ dependence of $T_c$ demonstrating 
  that the polarized quasiparticle injection 
  is insignificant. 
  Inset: Magnetization $m = n_\uparrow - n_\downarrow$ (circles) and 
  charge density $n = n_\uparrow + n_\downarrow $ (squares) profiles for 
  the $N=8$ trilayer with $J_{AF}=0.6 t$. 
  $n_\sigma = n_\sigma^{Cu (Mn)}$ in the cuprate (manganite) region 
  at $5 \le z \le 12$ ($z<5$ and $z>12$). 
  Results at $T=0 (T_c)$ are shown as open (filled) symbols. 
 }
\label{fig:Tcvsn_Mvsn}
\end{figure}

{\em Unconventional proximity effect.}---
Figure \ref{fig:Tcvsn_Mvsn} (main panel) shows $T_c$ as a function of $N$ for a trilayer in the P configuration. 
$T_c$ is strongly suppressed by reducing $N$, and SC is absent for $N=2$.  
Both the thickness at which SC disappears, and the length scale for a reduction in $T_c$ are very 
similar to the experimental results \cite{Sefrioui03}. 
For $J_{AF}=0$, $T_c$ remains relatively unchanged as a function of $N$.   
This demonstrates that the AF coupling plays a key role in the phenomena discussed here, 
while the injection of highly polarized quasiparticles is insignificant. 

The magnetization density $m$ profiles in Fig. \ref{fig:Tcvsn_Mvsn} (inset) 
show the origin of the characteristic length scale of the unconventional proximity effect. 
$m$ in the cuprate layer is 
more than 1 order of magnitude smaller than in the manganite layer, 
even for the Cu ions closest to the surface. 
In the absence of SC ($T = T_c$),  $m$ decays faster into the cuprate layer. 
Screening sets the length scale for the unconventional proximity effect. 
When a strong SC phase is present, 
it reduces $m$ near the surface, 
but the screening takes place in an oscillatory manner. 

\begin{figure}[tbp] 
\includegraphics[width=0.8\columnwidth,clip]{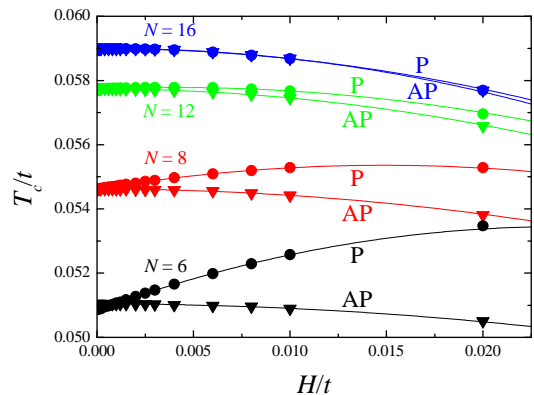} 
\caption{$T_c$ as a function of the applied magnetic field $H$ 
for different $N$. 
Circles (triangles) indicate $T_c$'s for the  P (AP) configuration  
(see Fig. \ref{fig:model}). 
Continuous lines are quadratic fits; 
the linear coefficients determine the behavior of $T_c (H)$ for the experimentally relevant fields.
A wide range of
magnetic fields is considered (note that $t
\approx$ 0.2-0.5 eV).  
}      
\label{fig:TcvsH} 
\end{figure}

{\em ISSE.}---
Next we consider the effect of an applied magnetic field in trilayer
systems, i.e., the ISSE. 
Figure \ref{fig:TcvsH} displays $T_c$ as a function of $H$
for the various $N$ indicated.  
The filled circles are numerical results, and the lines are quadratic
fits that will be discussed later. 
Changes in energy associated with the $H$'s are small, 
and it is expected that the relevant properties can be expanded in power series of the field. 
For the AP configuration, the system is symmetric against the inversion of $H$, 
and therefore the linear term in the expansion of $T_c$ is zero. 
For the P configuration, the linear term is always positive for the
different $N$ investigated.  This explains the surprisingly low field negative
magnetoresistance experimentally observed for trilayers in the P
configuration \cite{Nemes08}. 

A qualitative picture of these effects can be drawn from 
Figs. \ref{fig:model} and \ref{fig:Tcvsn_Mvsn}.
Without $H$, there are some small induced magnetic
moments in the cuprate region [Figs.~\ref{fig:model}(c) and ~\ref{fig:model}(d), Fig.~\ref{fig:Tcvsn_Mvsn} (inset)].
For the P configuration, applied fields compensate the effective 
field due to the AF coupling with manganites, especially at the exponential tail part [Fig.~\ref{fig:model}(e)].  
Thus, $T_c$ increases at low and moderate field regimes. 
In the AP configuration, effective and applied fields compensate only at one interface region 
but add up at the other.
The ISSE is therefore caused by the different effects of the external fields in the P and AP configurations. 
Experimental details, such as shape anisotropy, determine the switching fields 
and would induce rather nonlinear and hysteretic behavior.

For the $N=6$ trilayer with no applied field, 
$T_c$ in the P configuration is lower than in the AP configuration. 
The effective negative field  is negligible far enough into the
cuprate layer, but large near the interfaces. For small $N$, effective
fields from the different interfaces overlap. They
partially cancel in the AP configuration, 
while they add up in the P configuration. Therefore, $T_c$ is lower
for this last case. 
Nevertheless, nonzero $H$ produces a spectacular rise in $T_c$,  
as the interfacial effective field is very large. 
Even for $N=2$, $T_c$ becomes finite at $H > 0.035$.
The sensitivity of the thinnest layers to the magnetization orientation might contribute to  
the reported dependence of the SC properties on the manganite layer thickness \cite{Sefrioui03}. 
But, additional effects such as phase fluctuations not considered here might be important in this limit.

For $N=8$ and thicker trilayers, 
$T_c(H=0)$ for the P and AP configurations are the same
within our numerical precision. 
The effective fields corresponding to different interfaces do not overlap.
The P configuration has a higher $T_c$ under an applied field 
due to the partial cancellation of the external magnetic field and the effective field from the interface, giving rise to the ISSE.

\begin{figure}[tbp] 
\includegraphics[width=0.75\columnwidth,clip]{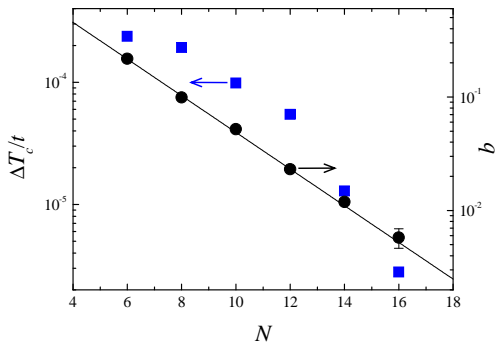} 
\caption{Change in the critical temperature $\Delta T_c = T^P_c - T^{AP}_c$ 
due to the different relative orientations of the magnetization of the manganite layers 
at $H=2 \times 10^{-3}t$ 
  as a function of $N$ (squares). 
  The exponential behavior is due to the linear dependence of $T_c^P$ ($H$). 
  For the fields accessible to experiments, this is the only relevant term, 
  and the linear coefficient $b$ also decays exponentially with $N$ (circles).}
\label{fig:bvsthickness}
\end{figure}

Finally, we discuss the change in $T_c$ due to the different alignment
of the FM layers $\Delta T_c=T_c^{P}-T_c^{AP}$ as a function of thickness. 
$ T_c^{P}$ and $T_c^{AP}$ change with an applied field 
(Fig. \ref{fig:TcvsH}).   
In experiments, the switching fields used to 
control the magnetic configuration 
depend on the details of the samples and the experimental setup. 
In Fig. \ref{fig:bvsthickness}, $\Delta T_c$ is defined as 
$T_c^{P}-T_c^{AP}$ at $H=2 \times 10^{-3}t$. 
With these switching field values, $\Delta T_c$ decays exponentially with $N$ 
(squares in Fig.\ref{fig:bvsthickness}), in agreement with Ref. \cite{Nemes08}. 
The choice of switching fields does not alter the exponential dependence of $\Delta T_c$ if these fields are small. 

For the fields accessible to experiments, $T^P_c$ increases linearly
with the field, 
while $T^{AP}_c$ only changes with the second power.
The zero field value of $T_c$ is equal for both configurations except for the thinnest layers. 
Thus, for the relevant switching fields $H_s$, $\Delta T_c = b H_s +{\cal O}(H_s^2)$, 
where $b$ is the (thickness dependent) linear coefficient in $T^P_c (H)$.
$b$ is plotted in Fig.~\ref{fig:bvsthickness} as a function of the cuprate layer thickness 
(circles). 
It is clear that the change in $b$ with respect to the cuprate thickness 
is the origin of the exponential decay in $\Delta T_c$, 
and that the switching field only appears as a vertical shift in the logarithmic scale of  Fig. \ref{fig:bvsthickness}.
The exponential decay length for the parameters used in this Letter is 
$l$=2.7 u.c., whic is smaller than 13 nm ($\approx$ 10 u.c.) found experimentally~\cite{Nemes08}. 
Fine-tuning of the parameters to make the experimental and theoretical results overlap is indeed possible, 
but vortices (not accounted for in the present mean field treatment) 
reduce the diamagnetic effect of the superconducting phase and are expected to 
increase the characteristic length of the {\it ISSE}. 
Our model provides a new explanation for this effect, 
reproduces the correct exponential behavior, and gives a qualitative agreement with experiments.  

To summarize, 
we investigated the proximity effect in manganite-cuprate systems 
with the AF coupling between the manganite and cuprate regions. 
This coupling induces {\it negative} spin polarization in the cuprate region, 
in contrast to the {\it positive} spin polarization expected from a simple charge transfer picture. 
The effect of the AF coupling was found to be much stronger than injecting spin-polarized quasiparticles. 
Various experimental anomalies are semiquantitatively reproduced based on our model calculations, 
such as the strong thickness dependence of $T_c$ in manganite-cuprate superlattices 
and the inverse spin-switch behavior of manganite-cuprate-manganite trilayers. 
As in this study, identifying proper interactions is crucial to understand 
the novel phenomena observed in many other oxide heterostructures. 

The authors thank J. Santamaria, M. Varela, and J. Chakhalian for their enlightening discussions. 
This work was supported by the NSF No. Grant DMR-0706020 (J.S.)
and by the Materials Sciences and Engineering Division, Office of
Basic Energy Sciences, the US DOE (S.O.).

\end{document}